\documentclass[12pt,notoc]{JHEP3}
\usepackage{amsfonts} %to get blackboard bold, bold etc
\usepackage{amsbsy} %likewise
\usepackage{amssymb} %to get more maths symbols eg \nsim
\usepackage{cite}

\newcommand{\C}[1]{{\mathcal #1}}
\newcommand{\beq}{\begin{equation}}
\newcommand{\eeq}{\end{equation}}
\newcommand{\bea}{\begin{eqnarray}}
\newcommand{\eea}{\end{eqnarray}}
\newcommand{\Tr}{{\hbox{Tr}\,}}
\newcommand{\comm}[2]{\left[#1,#2\right]}
\newcommand{\absval}[1]{{\left\vert#1\right\vert}}

\newcommand{\half}{{1\over 2}}

\def\Pf{\C P}
\def\const{ \hbox{\it{const}}}

\def\SYM{S_{\rm{YM}}}
\title{Adding a Myers Term to the IIB Matrix Model}

\author{Peter Austing\\ Department of Physics, University of Oxford \\
Theoretical Physics,\\
1 Keble Road,\\
 Oxford OX1 3NP, UK\\
E-mail: \email{p.austing@physics.ox.ac.uk}}

\author{John F. Wheater\\ Department of Physics, University of Oxford \\
Theoretical Physics,\\
1 Keble Road,\\
 Oxford OX1 3NP, UK\\
E-mail: \email{j.wheater@physics.ox.ac.uk}}

\preprint{hep-th/0310170\\OUTP/03-29P}
\abstract{We show that Yang-Mills matrix integrals remain convergent
when a Myers term is added, and stay in the same topological class as the original model. It is possible to add a supersymmetric
Myers term and this leaves the partition function invariant.}


\begin{document}

\section{Introduction}
The original motivation of
\cite{Myers:1999ps} for studying reduced Yang-Mills matrix models with
added cubic terms was 
to understand static $D0$ brane solutions in a 
constant RR background field. It was shown that separated branes
condense into a noncommutative space configuration, and this is known
as the Myers
effect. 

It is also interesting to add a Myers term to the action of
completely reduced
Yang-Mills matrix models. This gives 
an action which realises gauge theory on a noncommutative space
\cite{Iso:2001mg, Kitazawa:2002xj}, and would
therefore be an interesting deformation of the IIB matrix
model \cite{Ishibashi:1997xs}. In this article, we will make heavy use of the
existing analytic methods for arbitrary gauge groups \cite{Austing:2000rm,Austing:2001bd,Austing:2001pk,Austing:2001ib,Austing:2003cz} to show
that adding the Myers term \begin{it}is\end{it} a legitimate deformation of the original
matrix model.

As the author of \cite{Kitazawa:2002xj}
points out, the Myers term formally vanishes in the large $N$
limit. This means that it may act as a useful order parameter to study
a possible
dynamical reduction of the matrix model \cite{Aoki:1998vn} onto lower dimensional spaces
\cite{Iso:2001mg,Imai:2003vr,Imai:2003jb}. The possibility
of such
dynamical dimensional reduction of the model has been investigated since
early on
\cite{Aoki:1998vn,Ambjorn:2000bf,Ambjorn:2000dx,Ambjorn:2001xs,Bialas:2000gf,Burda:2000mn,Nishimura:2000wf,Nishimura:2001sx,Nishimura:2001sq,Anagnostopoulos:2001yb,Vernizzi:2002mu,Kawai:2002jk,Nishimura:2002va},
and for a recent review see \cite{Nishimura:2003rj}. In addition, one can consistently preserve half of the
supersymmetries of the matrix model while adding Myers type terms, and this is
one step in the MNS method \cite{Moore:1998et} for computing the SYMM
partition function (see also \cite{Krauth:1998yu,Krauth:1999qw,Krauth:1999rc,Krauth:2000bv,Staudacher:2000gx,Austing:2001ib,Pestun:2002rr}).

Addition of the Myers term takes the form
\beq
\SYM \rightarrow \SYM+ {i \over 3} \lambda \, f_{\mu \nu \rho}
\Tr \comm{X_\mu}{X_\nu} X_\rho
\eeq
where $f_{\mu \nu \rho}$ is an antisymmetric tensor.
If we assume that $\lambda$ is a real parameter, then we are led to a
puzzle. We have added unbounded cubic terms to the action,
and so one would guess that the partition function should
diverge. In the simplest example of a bosonic action with added Myers term
\beq
S= \Tr \comm{X_\mu}{X_\nu}\comm{X_\mu}{X_\nu}^\dagger +i \lambda \Tr X_1 \comm{X_2}{X_3},
\eeq
we can imagine fixing $\comm{X_2}{X_3}$ and taking $X_1$ to $\infty$ in such a way that the action would be unbounded from below. On the other hand, we could rewrite the trace as
\beq
i \Tr \comm{X_1}{X_2} X_3.
\eeq
Now it is the commutator $\comm{X_1}{X_2}$ which goes to 
$\infty$, and this can be overcome by the quadratic appearance of the
commutator in the Yang-Mills part of the action. However, it could be
that while $\comm{X_2}{X_3}$ itself was fixed, $X_3$ is going to $\infty$. The main purpose of this paper is to show that no matter how we take the matrices $X_1$, $X_2$ and $X_3$ to $\infty$, the large negative contribution to the action is always overcome in time to save convergence of the partition function.

Although it has been known for some time how to compute the $su(2)$ supersymmetric pure partition functions
(i.e. without Myers term) exactly \cite{Savvidy:1985gi,Smilga:1986jg,Smilga:1986nt,Yi:1997eg,Sethi:1997pa,Suyama:1998ig}, it was not realised until
\cite{Krauth:1998xh} that the bosonic integrals can also converge. It
was originally believed that integrating along the flat directions in
which the matrices commute would lead to divergence, while the Pfaffian
in the supersymmetric case could save the situation. However, the authors of
\cite{Krauth:1998xh} calculated the $su(2)$ bosonic partition
functions exactly, and found that they converge for $D\geq 5$ and
using careful numerical methods, found convergence for various other low
rank gauge groups
\cite{Krauth:1998xh,Krauth:1998yu,Krauth:1999qw,Krauth:1999rc,Krauth:2000bv,
Staudacher:2000gx}. In \cite{Austing:2001bd} we gave analytic
convergence criteria for the $su(N)$ bosonic models, and extended
these to the other gauge groups and
supersymmetric models in \cite{Austing:2001pk}. These methods have also
been used to study the large momentum behaviour of Polyakov lines
\cite{Austing:2003cz}, and we use them again here to obtain the results of this work.

In a recent article \cite{Tomino:2003hb}, D. Tomino obtains a finite result
for the $D=3$ $su(2)$ partition function with Myers term. Our analysis
does not apply to this case
since, while the integral without Myers term is formally zero, it is
absolutely divergent. As we shall see, even when the Myers term is large,
it is always small compared to the bosonic part of the action. This
seems to allow the same cancellations which saved the pure partition
function to also save the version with Myers term in this case.

In this note, we show that in fact the partition function with added
Myers term \begin{it}is\end{it} convergent as long as the pure
$\lambda=0$ partition function converges absolutely. We also show that adding a Myers term gives a true deformation of the pure Yang-Mills model and does not lead to a different topological class of models. This is clearly important if we are to use it as a probe of symmetry breaking in the pure model, and is not obvious; the supersymmetric mass terms added in the MNS method for example lead to a model which diverges when the masses  go to zero \cite{Moore:1998et,Austing:2001ib}. Finally, we write down
supersymmetric Myers terms and show that, in this case, the partition
function takes the same value as the $\lambda=0$ version, that is it is independent
of $\lambda$.

\section{Dealing with a Myers term}
We consider integrals of the form
\beq\label{2.1}
\C Z(\lambda) =\int 
\prod_{\mu=1}^D dX_\mu \,
\Pf (X_\mu )
\exp \left( -S_D(X) + i \lambda \epsilon_{\mu \nu \rho}
\Tr \comm{X_\mu}{X_\nu} X_\rho  \right) 
\eeq
where $S_D(X)= \Tr \comm{X_\mu}{X_\nu}\comm{X_\mu}{X_\nu}^\dagger$ is the bosonic action and $\Pf (X)$ is a Pfaffian generated by integrating out any fermions. We cover the supersymmetric cases, and also the bosonic case in which the Pfaffian is absent. We use the standard properties of Lie algebras which apply to Yang-Mills integrals and are described in for example \cite{Austing:2001pk}.

The problem is that terms of the form $i \lambda \Tr \comm{X_1}{X_2}
X_3$ are real and can be large and positive as well as negative. To
deal with this, we show that
\beq\label{3.1}
\absval{i \Tr \comm{X_1}{X_2}
X_3} \leq C_1  \, \absval{\Tr \comm{X_1}{X_2}^2
+\comm{X_2}{X_3}^2+\comm{X_3}{X_1}^2} + C_2
\eeq
that is, that the Myers term can be bounded by the bosonic part of the
action plus a constant.

Lets suppose that 
\beq 
\absval{X_1} \geq \absval{X_2}, \absval{X_3}
\eeq
and restrict to the region in which $\absval{X_1}>1$.
We can diagonalise $X_1$
\beq
X_1 \rightarrow \underline{x} \cdot H,
\eeq
where $H^i$ are the Cartan generators. It is convenient to use a basis in which $x_i= \underline{x} \cdot s_i$, where $\{s_i \}$ are the simple roots. Then by a Weyl transformation, we can assume \cite{Austing:2001pk}
\bea
x_1 , \cdots , x_r \geq 0 \label{pos}\\
x_1 \geq x_2 \geq \cdots \geq x_r, \label{order}
\eea
That is, first we choose our definition of positivity of roots so that
\ref{pos} is true, and then we relabel the simple roots so
that \ref{order} is true.

Then we can rewrite the Myers term as
\beq
i \lambda \Tr X_1 \comm{X_2}{X_3}
 = \lambda \sum_\alpha (\underline{x} \cdot \alpha) X_2^\alpha X_3 ^{-\alpha}
\eeq
where the sum is over all roots. Recall that every root can be written
$\alpha= \pm \sum_{i=1}^r n_i s_i$ where the $n_i$ are nonnegative integers.
Then we look at just one of the terms, $\alpha$, in the sum, for which
the root $\alpha$ contains $s_1$. Then since $\absval{X_1}\geq 1$ and
since $x_1$ is the biggest $x_i$, we have
\beq
\absval{(\underline{x} \cdot \alpha) X_2^\alpha X_3 ^{-\alpha}} \leq
\const (\underline{x} \cdot \alpha)^2 \absval{X_2^\alpha}^2
\label{3.7}
\eeq
where, without loss of generality, we have assumed that
$\absval{X_2^\alpha} \geq \absval{X_3^\alpha}$.

The rhs of \ref{3.7} is just one term in the expansion $\absval{\Tr
\comm{X_1}{X_2}^2} = \sum_\alpha (\underline{x} \cdot \alpha)^2 \absval{X_2^\alpha}^2$ and so we have the bound
\beq
\absval{\underline{x}\cdot \alpha X_2^\alpha X_3^{-\alpha}} \leq
\const \absval{\Tr \comm{X_1}{X_2}^2} + \const
\eeq
whenever $\alpha$ contains $s_1$.
Here we have included the additive constant to take care of the case
in which $\absval{X_1} \not> 1$.

At this stage, we have dealt with all the terms $\alpha$ containing
$s_1$, and so we have reduced the problem to that for the regularly
embedded subalgebra obtained by removing the simple root $s_1$. Then,
by induction, we have the result \ref{3.1}.

As usual, our strategy is to prove absolute convergence of the
partition function. So far, we have the bound
\beq
\absval{\exp\left( -S_D(X) +i  \epsilon_{\mu \nu \rho}
\Tr \comm{X_\mu}{X_\nu} X_\rho  \right) } \leq \exp\left
( -S_D(X) + C_1 S_D(X) + C_2 \right).
\eeq
The right hand side would only give a convergent integral if
$C_1<1$. However, the divergence in the integrals comes from the
region in which the radial variable $R= \sqrt{\Tr X_\mu X_\mu}$ goes
to infinity. Since the Myers term is cubic, while the bosonic part of
the action, $S_D$ is quartic, we can choose $C_1$ to be as small as we
like by looking at a region with $R$ large enough. We deduce that the
partition function
\beq\label{3.10}
\C Z(\lambda) =\int 
\prod_{\mu=1}^D dX_\mu \,
\Pf (X_\mu )
\exp \left( -S_D(X) + i \lambda \epsilon_{\mu \nu \rho}
\Tr \comm{X_\mu}{X_\nu} X_\rho  \right) 
\eeq
is convergent for any $\lambda$, as long as the pure Yang-Mills
integral $\C Z(0)$ is absolutely convergent.

It would be interesting to know whether the Myers term gives a continuous
deformation of the Yang-Mills integral, or whether it gives a
different toplogical class of model. In other words, we would like to
know whether the limit of $\C Z(\lambda)$ is $\C Z(0)$ when $\lambda
\rightarrow 0$. We can now answer this very simply. By Taylor's
theorem,
\bea
\C Z(\lambda)-\C Z(0) &= \lambda & \int 
\prod_{\mu=1}^D dX_\mu \, 
i 
\Tr\left( \comm{X_1}{X_2} X_3\right) \, \Pf (X_\mu ) \nonumber \\&& 
\;\;\; \exp \left( -S_D(X) + i \hat{\lambda} 
\Tr \comm{X_1}{X_2} X_3  \right)
\eea
for some $| \hat{\lambda} | < \absval{\lambda}$. Here we have chosen a
particular Myers term without losing any generality. We just need to check
that the integrals in the rhs are convergent, and as usual, the only
possible region of divergence is at large $R$. At large $R$, we can
bound the rhs by
\beq\label{3.12}
\lambda \int 
\prod_{\mu=1}^D dX_\mu \, 
S_D(X)  \, \absval{\Pf (X) }
\exp \left( -S_D(X) + \hat{\lambda} S_D(X)
  \right)
\eeq
up to a constant factor. Finally, we need to recall some details from
the proofs of convergence \cite{Austing:2001bd,Austing:2001pk}. We can
restrict integration to the region in which $S_D(X)<R^\epsilon$ where $\epsilon$ is
an arbitrarily small positive constant, since outside this region the
integrand decays exponentially. If we integrate out the angular
variables we obtain a bound
\beq
\int_1^\infty {dR \over R} R^{\epsilon} R^{-k_c(D)+ \epsilon'}
\eeq
where the $k_c$ are numbers, depending on the algebra as well as the
dimension $D$, which are calculated explicitly in
\cite{Austing:2001pk}, and $\epsilon'$ is another arbitrarily small,
positive constant. When the pure ($\lambda=0$) matrix model is
convergent, $k_c(D)$ must be positive, and so choosing $\epsilon$,
$\epsilon'$ small enough we find that \ref{3.12} is indeed
convergent. Finally then, we have
\beq
\absval{\C Z(\lambda) -\C Z(0)} < \const \, \lambda \rightarrow 0
\eeq
as $\lambda \rightarrow 0$.

\section{Supersymmetric Myers Term}
It is possible to add a Myers term to the supersymmetric models and
still preserve half of the supersymmetry \cite{Moore:1998et,
Austing:2001ib}. For example, for $D=4$, we can write the action
\bea
S &=& \Tr \left( (H + \half \comm{X_1}{X_2})^2 - {1 \over
4} \sum_{\mu>\nu}\comm{X_\mu}{X_\nu}^2 +i\lambda
\Tr(-X_3+iX_4)\comm{X_1}{X_2} \right.  \\
&&
\left.  - \epsilon_{ab} \eta_1 \comm{\psi_a}{X_b}
-\eta_a \half \comm{ (X_3+iX_4)}{\eta_a} -\psi_a \half \comm{
(-X_3+iX_4)}{\psi_a} + \eta_2 \comm{\psi_a}{X_a} \right). \nonumber
\eea
where $H$ is an auxilliary matrix, and the $\eta_a$ and $\psi_a$,
$a=1,2$ are the fermions. We have added the Myers term $i \lambda
\Tr(-X_3+iX_4)\comm{X_1}{X_2}$.

The action is $\delta$-exact, $S= \delta \Tr Q$, where
\beq\label{Q}
Q = \left( \eta_{1} [X_{1},X_{2}] +  \eta_{1} H + \half \psi_{a}
 [X_{a},X_3-iX_4] - {i \over 2} \eta_{2} [X_3 , X_4 ] \right)
\eeq
and
\beq\label{delta}
\begin{array}{lllllll}
\delta X_{a} & = & \psi_{a} \, , & & \delta \psi_{a} & = & \half [X_3 + i
X_4,X_{a}]
+i\lambda \, \epsilon_{ab} X_{b} \\ 
\delta \, X_3 & = & \eta_{2} \, , & & \delta \eta_{2} &=& {i \over 2}[X_4,X_3] \\
\delta \eta_{1} & = & H  \, , & &
 \delta
H & = & \half [X_3+iX_4,\eta_{1}] \\
\delta X_4 & = & i \eta_2
\end{array}
\eeq
and since $\delta$ squares to a gauge transformation plus a rotation,
one can quickly see that
$\delta S=0$. The same construction can also be made in $D=6$ and
$D=10$ \cite{Moore:1998et}, and one can check that, in $D=4$
for example, two of the four supercharges can be preserved in this way
\cite{Austing:2001ib}. The supercharge \ref{delta} represents a deformation of the pure ($\lambda=0$) version. It is not possible to generate a pure Myers term (ie without additional fermion masses) without deforming the supercharge. One can quickly see this since it would have to take the form $\delta \Tr P$ with $P$ a polynomial in the matrices \cite{Austing:2000rm}.

The results of the previous section show us that the partition function
\beq
\C Z(\lambda) = \int \exp(-S)
\eeq
is convergent, and that $\C Z(\lambda) \rightarrow \C Z(0)$ as
$\lambda \rightarrow 0$. With the above technology, one can now see
that $\C Z$ is independent of $\lambda$. This is a property
of rotation invariance rather than supersymmetry, but it will only work
for a Myers term of the given form which \begin{it}is\end{it}
imposed by supersymmetry.

First, since $\partial \C Z \over \partial \lambda$ exists and is
continuous for all $\lambda$, and ${\partial \C Z \over \partial
\overline{\lambda}}=0$, $\C Z(\lambda)$ is analytic. But, by a $2$D rotation
of the $X_3$, $X_4$ matrices, we have also\footnote{We are grateful to
M. Staudacher for originally pointing out this kind of phase argument.} 
\beq
\C Z(\lambda)= \C Z(e^{i \theta} \lambda).
\eeq
That is, $\C Z$ depends only on the magnitude of $\lambda$ and is
therefore independent of $\lambda$.

\acknowledgments{We are grateful to Graziano Vernizzi for valuable
discussion, and PA acknowledges an EPSRC fellowship.}
% \bibliography{../Bibliography/sources}

\begin{thebibliography}{10}

\bibitem{Myers:1999ps}
R.~C. Myers, {\it Dielectric-branes},  {\em JHEP} {\bf 12} (1999) 022,
  [\href{http://xxx.lanl.gov/abs/hep-th/9910053}{{\tt hep-th/9910053}}].

\bibitem{Iso:2001mg}
S.~Iso, Y.~Kimura, K.~Tanaka, and K.~Wakatsuki, {\it Noncommutative gauge
  theory on fuzzy sphere from matrix model},  {\em Nucl. Phys.} {\bf B604}
  (2001) 121--147, [\href{http://xxx.lanl.gov/abs/hep-th/0101102}{{\tt
  hep-th/0101102}}].

\bibitem{Kitazawa:2002xj}
Y.~Kitazawa, {\it Matrix models in homogeneous spaces},  {\em Nucl. Phys.} {\bf
  B642} (2002) 210--226, [\href{http://xxx.lanl.gov/abs/hep-th/0207115}{{\tt
  hep-th/0207115}}].

\bibitem{Ishibashi:1997xs}
N.~Ishibashi, H.~Kawai, Y.~Kitazawa, and A.~Tsuchiya, {\it {A large-N reduced
  model as superstring}},  {\em Nucl. Phys.} {\bf B498} (1997) 467--491,
  [\href{http://xxx.lanl.gov/abs/hep-th/9612115}{{\tt hep-th/9612115}}].

\bibitem{Austing:2000rm}
P.~Austing, {\it The cohomological supercharge},  {\em JHEP} {\bf 01} (2001)
  009, [\href{http://xxx.lanl.gov/abs/hep-th/0011211}{{\tt hep-th/0011211}}].

\bibitem{Austing:2001bd}
P.~Austing and J.~F. Wheater, {\it The convergence of Yang-Mills integrals},
  {\em JHEP} {\bf 02} (2001) 028,
  [\href{http://xxx.lanl.gov/abs/hep-th/0101071}{{\tt hep-th/0101071}}].

\bibitem{Austing:2001pk}
P.~Austing and J.~F. Wheater, {\it Convergent Yang-Mills matrix theories},
  {\em JHEP} {\bf 04} (2001) 019,
  [\href{http://xxx.lanl.gov/abs/hep-th/0103159}{{\tt hep-th/0103159}}].

\bibitem{Austing:2001ib}
P.~Austing, {\it {Yang-Mills matrix theory, PhD Thesis}},
  \href{http://xxx.lanl.gov/abs/hep-th/0108128}{{\tt hep-th/0108128}}.

\bibitem{Austing:2003cz}
P.~Austing, G.~Vernizzi, and J.~F. Wheater, {\it Polyakov lines in Yang-Mills
  matrix models},  {\em JHEP} {\bf 09} (2003) 023,
  [\href{http://xxx.lanl.gov/abs/hep-th/0309026}{{\tt hep-th/0309026}}].

\bibitem{Aoki:1998vn}
H.~Aoki, S.~Iso, H.~Kawai, Y.~Kitazawa, and T.~Tada, {\it Space-time structures
  from iib matrix model},  {\em Prog. Theor. Phys.} {\bf 99} (1998) 713--746,
  [\href{http://xxx.lanl.gov/abs/hep-th/9802085}{{\tt hep-th/9802085}}].

\bibitem{Imai:2003vr}
G.~U. A.~S. Imai, Takaaki AF~Tsukuba, Y.~Kitazawa, Y.~Takayama, and D.~Tomino,
  {\it Quantum corrections on fuzzy sphere},  {\em Nucl. Phys.} {\bf B665}
  (2003) 520--544, [\href{http://xxx.lanl.gov/abs/hep-th/0303120}{{\tt
  hep-th/0303120}}].

\bibitem{Imai:2003jb}
T.~Imai, Y.~Kitazawa, Y.~Takayama, and D.~Tomino, {\it Effective actions of
  matrix models on homogeneous spaces},
  \href{http://xxx.lanl.gov/abs/hep-th/0307007}{{\tt hep-th/0307007}}.

\bibitem{Ambjorn:2000bf}
J.~Ambjorn, K.~N. Anagnostopoulos, W.~Bietenholz, T.~Hotta, and J.~Nishimura,
  {\it Large N dynamics of dimensionally reduced 4d su(N) super Yang-Mills
  theory},  {\em JHEP} {\bf 07} (2000) 013,
  [\href{http://xxx.lanl.gov/abs/hep-th/0003208}{{\tt hep-th/0003208}}].

\bibitem{Ambjorn:2000dx}
J.~Ambjorn, K.~N. Anagnostopoulos, W.~Bietenholz, T.~Hotta, and J.~Nishimura,
  {\it Monte carlo studies of the iib matrix model at large N},  {\em JHEP}
  {\bf 07} (2000) 011, [\href{http://xxx.lanl.gov/abs/hep-th/0005147}{{\tt
  hep-th/0005147}}].

\bibitem{Ambjorn:2001xs}
J.~Ambjorn, K.~N. Anagnostopoulos, W.~Bietenholz, F.~Hofheinz, and
  J.~Nishimura, {\it On the spontaneous breakdown of lorentz symmetry in matrix
  models of superstrings},  {\em Phys. Rev.} {\bf D65} (2002) 086001,
  [\href{http://xxx.lanl.gov/abs/hep-th/0104260}{{\tt hep-th/0104260}}].

\bibitem{Bialas:2000gf}
P.~Bialas, Z.~Burda, B.~Petersson, and J.~Tabaczek, {\it Large N limit of the
  IKKT matrix model},  {\em Nucl. Phys.} {\bf B592} (2001) 391--407,
  [\href{http://xxx.lanl.gov/abs/hep-lat/0007013}{{\tt hep-lat/0007013}}].

\bibitem{Burda:2000mn}
Z.~Burda, B.~Petersson, and J.~Tabaczek, {\it Geometry of reduced
  supersymmetric 4d Yang-Mills integrals},
  \href{http://xxx.lanl.gov/abs/hep-lat/0012001}{{\tt hep-lat/0012001}}.

\bibitem{Nishimura:2000wf}
J.~Nishimura and G.~Vernizzi, {\it Brane world generated dynamically from
  string type iib matrices},  {\em Phys. Rev. Lett.} {\bf 85} (2000)
  4664--4667, [\href{http://xxx.lanl.gov/abs/hep-th/0007022}{{\tt
  hep-th/0007022}}].

\bibitem{Nishimura:2001sx}
J.~Nishimura and F.~Sugino, {\it Dynamical generation of four-dimensional
  space-time in the iib matrix model},  {\em JHEP} {\bf 05} (2002) 001,
  [\href{http://xxx.lanl.gov/abs/hep-th/0111102}{{\tt hep-th/0111102}}].

\bibitem{Nishimura:2001sq}
J.~Nishimura, {\it Exactly solvable matrix models for the dynamical generation
  of space-time in superstring theory},  {\em Phys. Rev.} {\bf D65} (2002)
  105012, [\href{http://xxx.lanl.gov/abs/hep-th/0108070}{{\tt
  hep-th/0108070}}].

\bibitem{Anagnostopoulos:2001yb}
K.~N. Anagnostopoulos and J.~Nishimura, {\it New approach to the complex-action
  problem and its application to a nonperturbative study of superstring
  theory},  {\em Phys. Rev.} {\bf D66} (2002) 106008,
  [\href{http://xxx.lanl.gov/abs/hep-th/0108041}{{\tt hep-th/0108041}}].

\bibitem{Vernizzi:2002mu}
G.~Vernizzi and J.~F. Wheater, {\it Rotational symmetry breaking in
  multi-matrix models},  {\em Phys. Rev.} {\bf D66} (2002) 085024,
  [\href{http://xxx.lanl.gov/abs/hep-th/0206226}{{\tt hep-th/0206226}}].

\bibitem{Kawai:2002jk}
H.~Kawai, S.~Kawamoto, T.~Kuroki, T.~Matsuo, and S.~Shinohara, {\it Mean field
  approximation of iib matrix model and emergence of four dimensional
  space-time},  {\em Nucl. Phys.} {\bf B647} (2002) 153--189,
  [\href{http://xxx.lanl.gov/abs/hep-th/0204240}{{\tt hep-th/0204240}}].

\bibitem{Nishimura:2002va}
J.~Nishimura, T.~Okubo, and F.~Sugino, {\it Convergent gaussian expansion
  method: demonstration in reduced Yang-Mills integrals},  {\em JHEP} {\bf 10}
  (2002) 043, [\href{http://xxx.lanl.gov/abs/hep-th/0205253}{{\tt
  hep-th/0205253}}].

\bibitem{Nishimura:2003rj}
J.~Nishimura, {\it Lattice superstring and noncommutative geometry},
  \href{http://xxx.lanl.gov/abs/hep-lat/0310019}{{\tt hep-lat/0310019}}.

\bibitem{Moore:1998et}
G.~Moore, N.~Nekrasov, and S.~Shatashvili, {\it D-particle bound states and
  generalized instantons},  {\em Commun. Math. Phys.} {\bf 209} (2000) 77,
  [\href{http://xxx.lanl.gov/abs/hep-th/9803265}{{\tt hep-th/9803265}}].

\bibitem{Krauth:1998yu}
W.~Krauth and M.~Staudacher, {\it {Finite Yang-Mills integrals}},  {\em Phys.
  Lett.} {\bf B435} (1998) 350,
  [\href{http://xxx.lanl.gov/abs/hep-th/9804199}{{\tt hep-th/9804199}}].

\bibitem{Krauth:1999qw}
W.~Krauth and M.~Staudacher, {\it Eigenvalue distributions in Yang-Mills
  integrals},  {\em Phys. Lett.} {\bf B453} (1999) 253--257,
  [\href{http://xxx.lanl.gov/abs/hep-th/9902113}{{\tt hep-th/9902113}}].

\bibitem{Krauth:1999rc}
W.~Krauth, J.~Plefka, and M.~Staudacher, {\it {Yang-Mills Integrals}},  {\em
  Class. Quant. Grav.} {\bf 17} (2000) 1171,
  [\href{http://xxx.lanl.gov/abs/hep-th/9911170}{{\tt hep-th/9911170}}].

\bibitem{Krauth:2000bv}
W.~Krauth and M.~Staudacher, {\it {Yang-Mills integrals for orthogonal,
  symplectic and exceptional groups}},  {\em Nucl. Phys.} {\bf B584} (2000)
  641, [\href{http://xxx.lanl.gov/abs/hep-th/0004076}{{\tt hep-th/0004076}}].

\bibitem{Staudacher:2000gx}
M.~Staudacher, {\it {Bulk Witten indices and the number of normalizable ground
  states in supersymmetric quantum mechanics of orthogonal, symplectic and
  exceptional groups}},  {\em Phys. Lett.} {\bf B488} (2000) 194,
  [\href{http://xxx.lanl.gov/abs/hep-th/0006234}{{\tt hep-th/0006234}}].

\bibitem{Pestun:2002rr}
V.~Pestun, {\it N = 4 SYM matrix integrals for almost all simple gauge groups
  (except E(7) and E(8))},  {\em JHEP} {\bf 09} (2002) 012,
  [\href{http://xxx.lanl.gov/abs/hep-th/0206069}{{\tt hep-th/0206069}}].

\bibitem{Savvidy:1985gi}
G.~K. Savvidy, {\it Yang-Mills quantum mechanics},  {\em Phys. Lett.} {\bf
  B159} (1985) 325.

\bibitem{Smilga:1986jg}
A.~V. Smilga, {\it {Witten index calculation in supersymmetric gauge theory}},
  {\em Nucl. Phys.} {\bf B266} (1986) 45--57.

\bibitem{Smilga:1986nt}
A.~V. Smilga, {\it Calculation of the Witten index in extended supersymmetric
  Yang-Mills theory. (in Russian)},  {\em Yad. Fiz.} {\bf 43} (1986) 215--218.

\bibitem{Yi:1997eg}
P.~Yi, {\it {Witten index and threshold bound states of D-branes}},  {\em Nucl.
  Phys.} {\bf B505} (1997) 307,
  [\href{http://xxx.lanl.gov/abs/hep-th/9704098}{{\tt hep-th/9704098}}].

\bibitem{Sethi:1997pa}
S.~Sethi and M.~Stern, {\it {D-brane bound states redux}},  {\em Commun. Math.
  Phys.} {\bf 194} (1998) 675,
  [\href{http://xxx.lanl.gov/abs/hep-th/9705046}{{\tt hep-th/9705046}}].

\bibitem{Suyama:1998ig}
T.~Suyama and A.~Tsuchiya, {\it Exact results in n(c) = 2 iib matrix model},
  {\em Prog. Theor. Phys.} {\bf 99} (1998) 321--325,
  [\href{http://xxx.lanl.gov/abs/hep-th/9711073}{{\tt hep-th/9711073}}].

\bibitem{Krauth:1998xh}
W.~Krauth, H.~Nicolai, and M.~Staudacher, {\it {Monte Carlo approach to
  M-theory}},  {\em Phys. Lett.} {\bf B431} (1998) 31--41,
  [\href{http://xxx.lanl.gov/abs/hep-th/9803117}{{\tt hep-th/9803117}}].

\bibitem{Tomino:2003hb}
D.~Tomino, {\it N=2 3d-matrix integral with Myers term},
  \href{http://xxx.lanl.gov/abs/hep-th/0309264}{{\tt hep-th/0309264}}.

\end{thebibliography}
\providecommand{\href}[2]{#2}\begingroup\raggedright\endgroup

\end{document}